\documentclass{article}
\usepackage{cite}
\usepackage{wrapfig}
\usepackage{graphicx}
\usepackage{amssymb}
\usepackage{amsfonts}
\usepackage{amsmath}
\usepackage{longtable}
\usepackage{rotating}
\usepackage{lscape}
\usepackage{epsfig}
\usepackage{multirow}

\begin{document}
\title{Hadron production measurements at NA61/SHINE for precise determination of accelerator neutrino fluxes}
\date{}
\author{S.\,Ilieva$^{a,b,}$\\
$^{a}$\,for the NA61/SHINE collaboration\\
$^{b}$\,Faculty of Physics, Sofia University "St. Kliment Ohridski",\\ Blvd "James Bourchier" 5, 1164 Sofia, Bulgaria }
\maketitle

\begin{abstract}
The total systematic uncertainty of the neutrino flux in accelerator-based neutrino experiments is dominated by the Monte Carlo modeling of hadronic interactions. Direct hadron production measurements for T2K and Fermilab neutrino experiments, MINER$\mathrm{\nu}$A, NO$\mathrm{\nu}$A and DUNE, are being performed at the NA61/SHINE spectrometer at CERN’s Super Proton Synchrotron. Crucial for improving neutrino flux predictions, hadron yields, inelastic and production cross sections are obtained at NA61/SHINE where interactions of various hadron beams with thin and thick (replica) targets are reproduced. Recently obtained results will be reported. An extension of NA61/SHINE’s program of hadron production measurements for neutrino experiments is planned beyond 2020.
\end{abstract}
\vspace*{6pt}

\noindent

\label{sec:intro}
\section{Introduction}
In long-baseline neutrino oscillation experiments direct hadron production measurements are crucial for reducing the uncertainty of the neutrino flux prediction at both near and far detectors. Dedicated hadron production measurements provide interaction cross sections and double differential particle yields. These results are in turn used to re-weight simulations of the neutrino production at a given neutrino beamline. Since 2007, direct hadron production measurements for J-PARC, NuMI and the future LBNF neutrino beamlines have been carried out at the NA61/SHINE experiment. 

\label{sec:na61}
\section{The NA61/SHINE experiment}
\begin{figure}[th]
\begin{center}
\includegraphics[width=110mm,height=56mm]{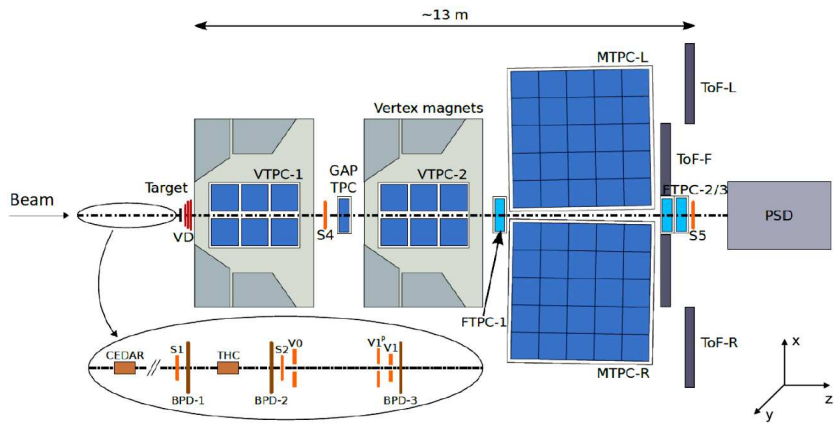}
\vspace{-3mm}
\caption{The NA61/SHINE detector setup. Upstream of the target are the scintillators (S1, S2, V0, $\mathrm{V1}^{p}$, V1), the Cherenkov detectors (CEDAR, THC) and the proportional chambers - Beam Position Detectors (BPD-1, BPD-2, BPD-3). Downstream of the target are the Vertex Detector (VD), the Vertex TPC 1 and 2 (VTPC-1 \& VTPC-2), the scintillators (S4, S5), the Gap TPC (GTPC), the Main TPC Left and Right (MTPC-L \& MTPC-R), the Forward TPCs (FTPC-1, FTPC-2 \& FTPC-3), the Time-of-Flight wall Left, Right and Forward (TOF-L \& TOF-R \& TOF-F) and the Projectile Spectator Detector (PSD).}
\label{fig1}
\end{center}
\vspace{-5mm}
\end{figure}

The NA61/SHINE Heavy Ion and Neutrino Experiment is a fixed-target experiment at the CERN's Super Proton Synchrotron (SPS)\cite{na61_det}. Situated on the H2 beamline, the experiment operates with hadron and ion beams in wide energy ranges: 13~GeV/c - 400~GeV/c for hadrons and  13A~GeV/c - 158A~GeV/c for ions. At the NA61/SHINE spectrometer hadron-proton, hadron-nucleus and nucleus-nucleus collisions are being examined. The detector setup is presented in Fig.~\ref{fig1}. Position of the incoming beam particles is determined by a set of three proportional chambers called Beam Position Detectors (BPD). Trigger is provided by multiple scintillators and Cherenkov detectors. Further downstream from the target is the silicon vertex detector (VD) that is used for precise vertex determination and allows for the measurement of $D^{0}$ mesons. The NA61/SHINE detector facilitates charged particle momentum measurements and particle identification by eight Time Projection Chambers(TPC) and three Time-of-Flight(ToF) detectors. Momentum determination is possible since three of the TPCs are inside a magnetic field that is provided by two superconducting dipole magnets. Maximal total bending power is 9 T\,m. Three of the TPCs, Forward TPCs or FTPCs, were installed in 2017 and have improved forward acceptance of the spectrometer extending it to the 100~GeV/c momenta  range \cite{ftpc}. Furthest from the target is a forward hadron calorimeter, called Projectile Spectator Detector (PSD), that measures the projectile spectator energy in heavy-ion collisions.\\
The NA61/SHINE experiment has a broad research program featuring strong-interaction physics, cosmic ray physics and neutrino physics. The experiment's neutrino program comprises of precise measurements of hadron production in various reactions and is set to improve the prediction of the neutrino flux at T2K and Fermilab neutrino experiments, MINER$\mathrm{\nu}$A, NO$\mathrm{\nu}$A and DUNE.

\label{sec:motivation}
\section{Necessity for direct hadron production measurements}
Accelerator neutrino beams are initiated by a high-energy hadron beam, typically a proton beam, that hits a thick target of light material producing secondary hadrons that further decay to neutrinos. Secondary hadrons can re-interact inside the target or with the beamline materials giving rise to additional neutrino-yielding particles. Modeling the above-mentioned hadron interactions is the leading source of systematic uncertainty of the neutrino flux prediction as different Monte Carlo models provide different estimates of particle yields and interaction cross sections. Moreover, hadrons of different type and energy contribute to different parts of the neutrino energy spectrum \cite{T2K_flux}. Neutrinos coming from the decay of hadrons produced in interactions of secondary nucleons inside the target dominate the low-energy range of the neutrino spectrum - below a few GeV. On the other hand, constraining kaon productions and so neutrinos produced in kaon decays reduces the uncertainty of the neutrino flux at energies larger than a few GeV. At long-baseline neutrino experiments, precise neutrino flux prediction is key for the prediction of neutrino interaction rates at the near and the far detector. As such, precise flux prediction is necessary for both neutrino oscillation and neutrino cross-section measurements. To scale simulation results and consequently better constrain the neutrino beam, external hadron production measurements are needed.

\label{sec:HadProd}
\section{Reference hadron production measurements}
The NA61/SHINE collaboration uses the following classification of nuclear interactions. Elastic processes are those in which no new particles are produced. Inelastic processes lead to the disintegration of the target nucleus and are divided into two subsets - production and quasi-elastic interactions. In production processes new hadrons are produced in the final state. In quasi-elastic processes the nucleus fragments following the interaction, but no new hadrons are produced.\\
Hadron production measurements at NA61/SHINE are made using thin and thick (replica) targets. Thin targets are a few centimeters long (a few $\%$ of nuclear interaction length, \textit{$\lambda_{i}$}), while replica targets are tens of centimeters in length (a few \textit{$\lambda_{i}$}). Primary hadron interactions in the target and the elements of the beamline are reproduced using thin targets of various materials, including carbon, beryllium, aluminum, and beams of protons, pions, and kaons. Such measurements give estimates of inelastic and production cross sections as well as differential cross sections and differential particle yields. The cross-section data are used to re-weight hadron absorption probabilities in the simulation of the neutrino production chain.\\
Primary and secondary interactions inside the target are simultaneously studied using replica targets and proton beams. Those measurements aim to provide differential hadron yields on target surface and a measure of the production cross section, based on the beam attenuation inside the target. The differential hadron yields are calculated in terms of momentum, \textit{p}, polar angle, \textit{$\theta$}, and target longitudinal, \textit{z}, bins. These data are used to re-weight the differential particle yields obtained with Monte Carlo for the same (\textit{p},~\textit{$\theta$},~\textit{z}) bin . 

\label{sec:results_p}
\section{Hadron production measurements in $\mathrm{p}$ interactions}
\begin{figure}[h]
\begin{center}
\includegraphics[width=110mm,height=56mm]{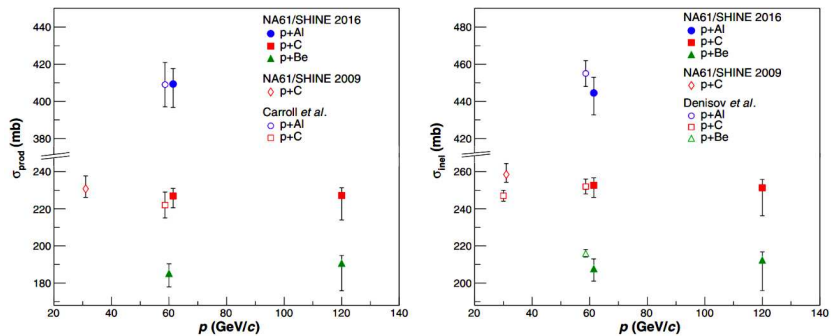}
\vspace{-3mm}
\caption{Summary of NA61/SHINE thin-target production(on the left) and inelastic(on the right) cross-section measurements. Results by Carroll \textit{et al.}~\cite{Carroll} and Denisov \textit{et al.}~\cite{Denisov} are also given for reference. For these measurements proton beams at 31, 60 and 120 GeV/c momenta and thin targets of carbon, aluminum and beryllium have been used.}
\label{fig2}
\end{center}
\vspace{-5mm}
\end{figure}

\begin{figure}[h]
\begin{center}
\includegraphics[width=110mm,height=56mm]{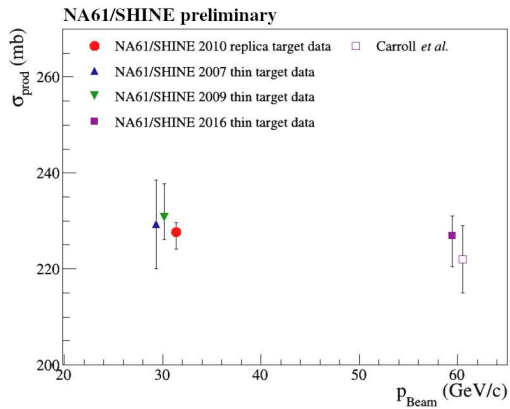}
\vspace{-3mm}
\caption{Production cross-section measurements for $\mathrm{p}$ + C interactions at different beam momenta. NA61/SHINE replica-target result at 31 GeV/c is marked with a circle. It is shown alongside NA61/SHINE thin-target results at 31 and 60 GeV/c and the production cross-section measurement by Carroll \textit{et al.}~\cite{Carroll} at 60 GeV/c.}
\label{fig3}
\end{center}
\vspace{-5mm}
\end{figure}

Latest results of the NA61/SHINE neutrino program in studies of proton interactions include both thin and replica target measurements. In particular, production and inelastic cross sections in $\mathrm{p}$ + C, $\mathrm{p}$ + Be, and $\mathrm{p}$ + Al at 60 GeV/c and $\mathrm{p}$ + C and $\mathrm{p}$ + Be at 120 GeV/c were estimated using thin targets \cite{na61_nagai}. These results are summarized in Fig.~\ref{fig2}, where for reference other measurements by Carroll \textit{et al.} \cite{Carroll} and by Denisov \textit{et al.} \cite{Denisov} are also given. The uncertainty of these NA61/SHINE measurements is dominated by the Monte Carlo model dependency, which is introduced through MC-based corrections for elastic and quasi-elastic scattering. The cross-section data for protons at 120 GeV/c beam momentum are the first ones in this high-energy domain.
\\An independent production cross-section measurement was performed with the T2K replica target - a 90-cm-long graphite cylinder with a 1.3-cm radius of the base. A proton beam at 31 GeV/c momentum was used. The cross-section result was directly extracted from the beam survival probability, $\textit{P}_\mathrm{surv}$, as per
\begin{equation}
    P_\mathrm{surv}=e^{-L n \sigma_\mathrm{prod}},
\end{equation}
where \textit{L} is the length of the target, \textit{n} is the number density of target nuclei and \textit{$\sigma$}$_\mathrm{prod}$ is the production cross section. In such a way, the model dependency is minimized and the precision of the final result is improved compared to previous NA61/SHINE thin-target measurements, as shown in Fig.~\ref{fig3}.
\\ NA61/SHINE has also performed measurements of double differential hadron yields on the surface of the T2K replica target. Charged pion yields were obtained following data-taking in 2009 \cite{t2k_2009yields}, while proton, charged pion and kaon yields \cite{t2k_2010yields} were estimated using data from the year 2010 run. Implementation of these measurements into the T2K neutrino flux prediction reduced the total flux uncertainty to about 5$\%$ at the peak beam energy of 0.6 GeV~\cite{na61_spsc}.\\
In 2018, NA61/SHINE took data with the NuMI replica target - a 120-cm-long sequence of graphite fins. A proton beam at 120 GeV/c momentum was used. These data are currently being calibrated.

\label{sec:results_pi}
\section{Hadron production measurements in $\mathrm{\pi^{+}}$ interactions}
\begin{figure}[h]
\begin{center}
\includegraphics[width=110mm,height=56mm]{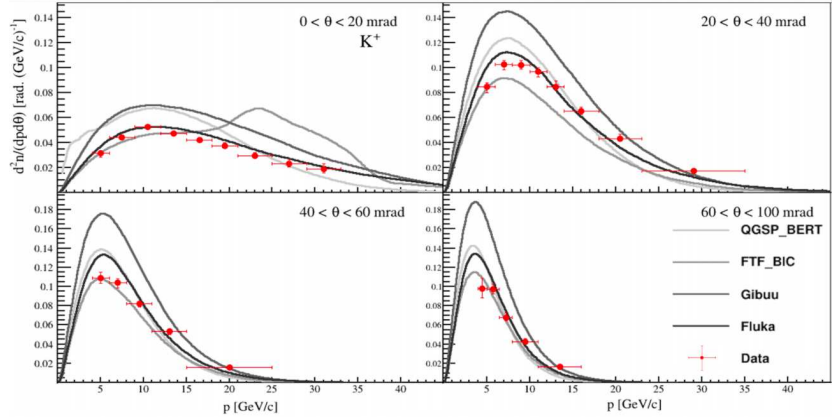}
\vspace{-3mm}
\caption{$\mathrm{K^{+}}$ double differential yields from $\mathrm{\pi^{+}}$ + C at 60 GeV/c interactions. Each plot corresponds to a given polar angle, \textit{$\theta$}, range. Data points are shown alongside predictions of the GEANT4~\cite{Geant4} physics lists: QGSP\textunderscore BERT and FTF\textunderscore BIC as well as GiBUU2019~\cite{GiBUU2019} and FLUKA2011~\cite{FLUKA} generators. The data error bars represent total uncertainties except for the normalization uncertainty. }
\label{fig4}
\end{center}
\vspace{-5mm}
\end{figure}

Positive pion interactions on thin carbon and beryllium targets were recently studied at NA61/SHINE \cite{na61_johnson}. This resulted in the first measurement of production cross section in $\mathrm{\pi^{+}}$ + Be at 60 GeV/c incident beam energy and a broad study of charged and neutral hadron production - $\mathrm{\pi^{\pm}}$, $\mathrm{K^{\pm}}$, $\mathrm{p}$, $\mathrm{K^{0}_{s}}$, $\mathrm{\Lambda}$. An example of the obtained double differential kaon yields is given in Fig.~\ref{fig4}, where data and MC estimates with different simulation generators are shown. In some (\textit{p}, \textit{$\theta$}) bins, Monte Carlo results significantly differ from the data. This stresses the need for direct hadron production measurements. These results are expected to reduce the uncertainties associated with secondary interactions of pions in the carbon targets and the beryllium elements of the NuMI and the future LBNF beamlines.
\newline
Similarly, positive pion interactions on a thin aluminum target were constrained with a direct measurement at NA61/SHINE \cite{fermi_2015}. Inelastic and production cross sections were obtained for $\mathrm{\pi^{+}}$ beams at 31 and at 60 GeV/c momenta on thin aluminum and carbon targets. In addition, same results were obtained for $\mathrm{K^{+}}$ beams at 60 GeV/c and thin carbon and aluminum targets. 

\label{sec:future}
\section{Future plans}
During CERN's Long Shutdown 2, the NA61/SHINE collaboration is working on several detector upgrades. For the experiment's future neutrino program, the following changes are particularly important. A replacement of the TPC readout electronics, together with new trigger and data acquisition system, will increase data rate to 1 kHz\footnote{current readout rate is $\sim$ 100 Hz} and will reduce statistical uncertainties. New ToF detectors with Multigap Resistive Plate Chamber (MRPC) technology will improve timing resolution. Dedicated research of the feasibility and the construction of a Very Low Energy tertiary beamline (beam momenta in range 1-20 GeV/c) is under action as well. Such a line will allow for studies of the low-energy interactions of protons, pions and kaons that are still unconstrained by direct measurements. These data are needed to further lower down the hadron interaction uncertainty of the neutrino flux predictions at long-baseline neutrino experiments.\\
NA61/SHINE will resume data taking in 2021 with a measurement of 31 GeV/c protons and the T2K replica target. The goal of this additional replica-target run is to increase the statistics for high-energy kaon production. Later on, in 2022-2024, more hadron production measurements are planned with kaon and proton beams on thin targets and with a proton beam at 120 GeV/c on the LBNF/DUNE prototype target. After Long Shutdown 3, plans for measurements of hadron yields off of LBNF and Hyper-Kamiokande replica targets are being considered.

\label{sec:summary}
\section{Summary}
Precise hadron production measurements are crucial for reducing the leading systematic uncertainty of the neutrino flux prediction at accelerator-based neutrino experiments. Thin-target and replica-target measurements by the NA61/SHINE experiment have already led to a significant reduction to $\sim$5$\%$ for the T2K neutrino flux prediction at its peak energy. A broad data-taking program for Fermilab neutrino experiments was carried out and several important analysis results have already been released. NA61/SHINE is currently working on a large detector upgrade. A possibility to construct a low-energy tertiary beamline is being studied. Resuming in 2021, NA61/SHINE plans for new hadron production measurements with both thin and replica targets, adhering to the needs for more precise flux predictions at current and future long-baseline neutrino experiments.\\ 

\section*{Acknowledgments}
    This work is supported by the Bulgarian National Science Fund (grants DN08/11 and DCOST01/8) and the Bulgarian Nuclear Regulatory Agency and the Joint Institute for Nuclear Research, Dubna according to bilateral contract No. 4799-1-18/20.
\bibliographystyle{ieeetr}
\bibliography{references}

\end{document}